\documentclass{PoS}
\usepackage{amsmath,graphicx}
\newcommand{\bo}{\raise-1mm\hbox{\Large$\Box$}}
\newcommand{\tmin}{t_{\text{min}}}

\PoS{PoS(LAT2005)029}

\title{Charmonium spectrum on dynamical anisotropic lattices}
\ShortTitle{Charmonium Spectrum}
\author{\speaker{K. Jimmy Juge}\\
Trinity College, Dublin, Ireland\\
E-mail: \email{juge@maths.tcd.ie}}
\author{Alan \'O Cais\\Trinity College, Dublin, Ireland\\
E-mail: \email{alanc@maths.tcd.ie}}
\author{Mehmet B. Oktay\thanks{Based on talks by KJJ and MBO.}\\Trinity College, Dublin, Ireland\\
E-mail: \email{oktay@maths.tcd.ie}}
\author{Mike J. Peardon\\Trinity College, Dublin, Ireland\\
E-mail: \email{mjp@maths.tcd.ie}}
\author{Sin\'ead M. Ryan\\Trinity College, Dublin, Ireland\\
E-mail: \email{ryan@maths.tcd.ie}}
\abstract{We present a first study of the charmonium spectrum on $N_f=2$ dynamical, anisotropic lattices. We take advantage of all-to-all quark propagators to build spatially extended interpolating operators to increase the overlap with states not easily accessible with point propagators such as radially excited states of $\eta_c$, $\psi$ and $\chi_c$, D-waves and hybrid states.}

\FullConference{
    XXIIIrd International Symposium on Lattice Field Theory\\
                 25-30 July 2005\\
                 Trinity College, Dublin, Ireland}

\begin{document}

\section{Introduction}
Many new states have been observed in the charmonium sector in the past few years. The missing spin singlet states as well as new, yet to be explained states such as the X(3872)\cite{Choi:2003ue}, Y(4260)\cite{Aubert:2005rm}, Y(3940)\cite{Abe:2004zs} and others have presented a new challenge for the theoretical particle physics community.

Simulating the charmonium spectrum in lattice QCD has long been
problematic due to the fact that the mass is too heavy for a simple
relativistic action on coarse lattices and too light to expand in
inverse powers of the quark mass. There are various quenched studies
with improved actions which attempt to reduce lattice artifacts
\cite{El-Khadra:1992ir,Okamoto:2001jb} but very little has been done
on dynamical configurations \cite{diPierro:2003bu}.

In this study, we use a new dynamical, anisotropic action which has
the advantage of being fully relativistic and yet has small
mass-dependent discretization errors. All-to-all propagators
\cite{Foley:2005ac} are utilized to make full use of the
configurations and to construct extended operators for better overlap
with the higher charmonium states.

\section{Dynamical anisotropic action}

The $N_f=2$ anisotropic quark action \cite{Foley:2004jf} is as follows:
\begin{equation}
S_q=\bar\psi\left(\gamma_0\nabla_0+\sum_i\mu_r\gamma_i\nabla_i\left(1-\frac{1}{\xi_q a_s^2}\Delta_i\right)-\frac{ra_t}{2}\Delta_{i0}+sa_s^3\sum_i\Delta_i^2+m_0\right)\psi\,,\label{eq:qaction}
\end{equation}
where the links are fattened using stout links
\cite{Morningstar:2003gk} which maximize the plaquette.  The target renormalized anisotropy, $\xi$, is 6. The nonperturbative tuning of the action has been presented elsewhere at this conference \cite{Morrin:2005zq}. The sea quark mass was set near the strange quark in this first study although there are indications that there are no major problems simulating at lighter sea quark masses. The gauge action is the ``two-plaquette'' action \cite{Morningstar:1999dh} which was designed to reduce the cutoff effects for the scalar glueball on coarse anisotropic lattices. We use the same anisotropic fermion action for both light sea quarks and heavy valence quarks. In a quenched study \cite{Foley:2004jf}, it was found that the same bare anisotropy can be used for a large range of quark masses. For the heavy valence quark, we have used both stout links and original links in this study. 

\section{Operators}

We use a variational basis of operators, many of which are extended in
space, to maximize the overlap with the higher charmonium
states. All-to-all propagators which do not introduce large variance
are crucial for this construction. We use the ``dilution'' method of
Ref.~\cite{Foley:2005ac} without eigenvectors for the charm quark
propagators. Several dilution schemes were tested on a small number of
configurations. We present results which use
``time+colour+space-even-odd'' dilution.  Preliminary investigations
indicate that even higher dilutions may further reduce the errors;
this is work in progress.

The S-wave charmonium states only require different smearings of the quark to project out radially excited states. However, we have included a hybrid $1^{--}$ operator which contains a gluonic excitation in the form of a chromomagnetic and ``staple'' field. The operators used in this study are summarized in Table~\ref{table:ops}. Many of the operators used here resemble the ones used in Ref.~\cite{Lacock:1996vy}. In addition to these basic operators, we have used two different smearings for the quark fields to build a larger variational basis. The variational optimization was performed at the largest timeslice possible (with the metric timeslice fixed at $t=1$) where the results have become independent of this choice. We note that the $\eta_c$ and $\chi_c$ states (especially the $0^{++}$) required a large $t$ for the results to become stable. This may be an indication that better operators can improve the situation for these channels significantly.

For the exotic channel, $1^{-+}$, we have used both the staple and
chromomagnetic field operator since we know that the staple operator
has a large overlap with the first excited gluonic state in the static
$q\bar{q}$ system \cite{Juge:2002br}. The chromomagnetic field
operator has the advantage of being an easy operator to construct as
it is an operator at a single point. We have found that there is very
little difference between the two operators, contrary to the
non-relativistic quark model which suggests that the quarks are in a
relative P-wave. This may be due to the fact that the quarks have been
smeared or that both operators have a high level of excited state
contamination from multi-particle states.

\begin{table}
\begin{center}
\begin{tabular}{cc}
\hline
 Particle & Operator \\ \hline\hline
$1^{--}$ & $\vec{\gamma}$, $\gamma_5\vec{u}$, $\gamma_5\vec{B}$\\
$0^{-+}$ & $\gamma_5$, $\gamma_5(s_1+s_2+s_3)$ \\
$0^{++}$ & $\vec{\gamma}\cdot\vec{p}$ \\
$1^{++}$ & $\vec{\gamma}\times \vec{p}$ \\
$2^{++}$ & $\gamma_kp_i+\gamma_ip_k$ \\
$1^{+-}$ & $\gamma_5\vec{p}$ \\
$1^{-+}$ & $\vec{\gamma}\times\vec{u}$, $\vec{\gamma}\times\vec{B}$ \\
$2^{-+}$ & $\gamma_5(s_1-s_3),\gamma_5(2s_3-s_1-s_3)$ \\
$3^{--}$ & $\vec{\gamma}\cdot\vec{t}$ \\\hline
\end{tabular}
\end{center}
\caption{Some operators used to project out various higher lying
  states. The notation for the gluonic paths is that of Ref.~\cite{Lacock:1996vy}.}
\label{table:ops}
\end{table}

\section{Analysis}

\subsection{Fitting scheme}
The use of time-diluted all-to-all propagators makes it difficult to
identify the plateau region (and judge its quality) from effective
mass plots. This is due to the random noise introduced at each
timeslice which makes locally measured quantities such as the
effective mass fluctuate more than they would with point
propagators. These ``local fluctuations'' do not affect exponential
fits as long as a reasonable range of timeslices are included
since these fits capture the long range exponential decay of the
correlation function. (This feature will disappear once the gauge
noise limit is reached in the dilution method. We already see a
noticeable decrease of these effects at the higher dilution results
that are presented here.) We therefore show instead ``$\tmin$'' plots
where fitted values of the mass are shown against the minimum
timeslice included in the single exponential fits. The largest
timeslice to include in the fit was taken to be halfway across the
lattice, if the fluctuations were not dominating the signal. We show a
typical example of an effective mass ($1^{++}$) and its corresponding
$\tmin$ plot on the left. Our criteria for selecting a fit is
based on the $\chi^2$, fit range and the fit quality Q. We demand
(although there could have been many other choices for the criteria
for a ``good fit'') that $\chi^2/N_{df}$ be less than 2, the
quality of the fit be larger than $0.2$ and that there is a range of
$\tmin$ where these fit values are stable.

\begin{figure}[ht]
\begin{center}
\includegraphics[width=2.8in]{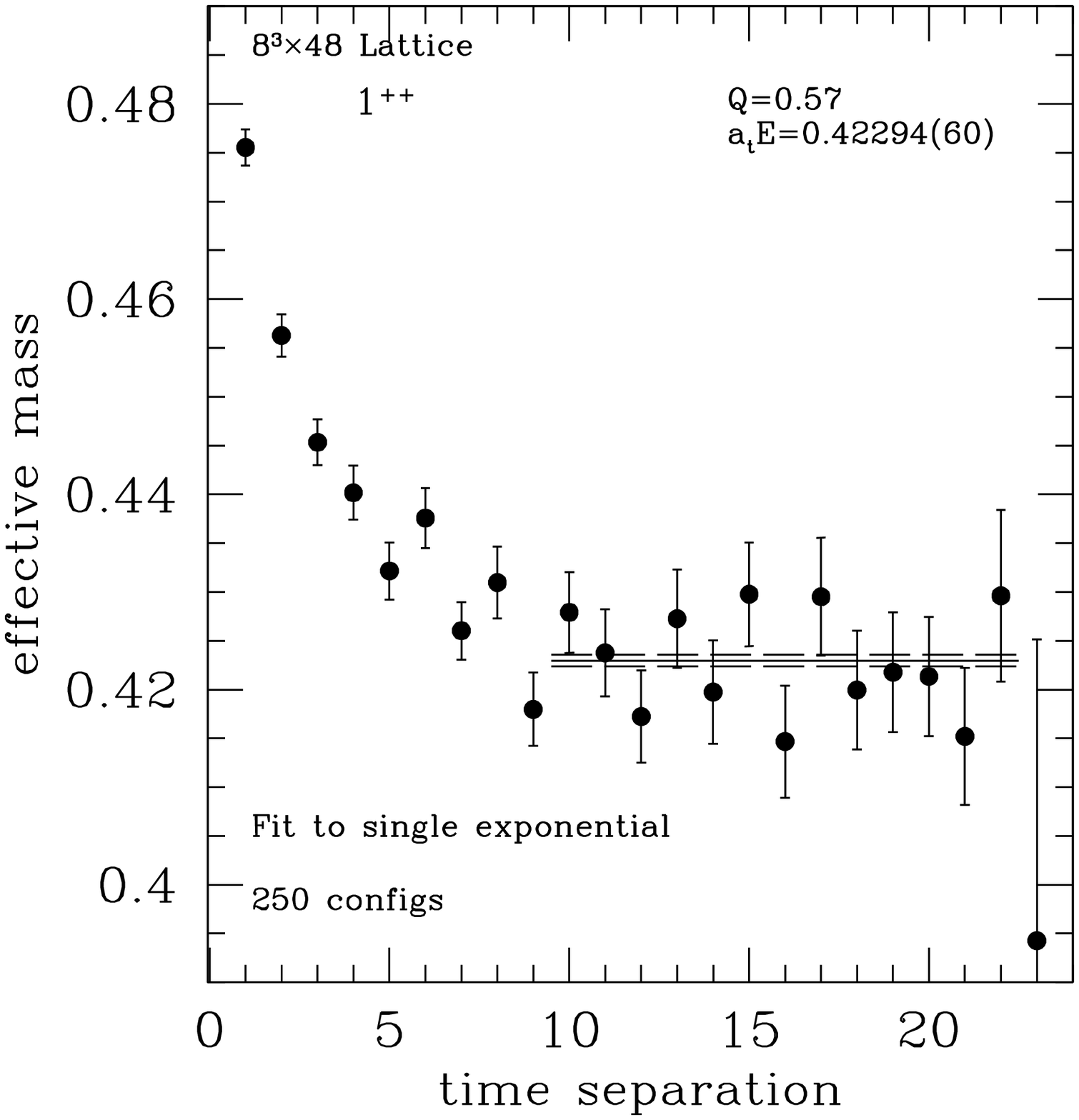}
\includegraphics[width=2.8in]{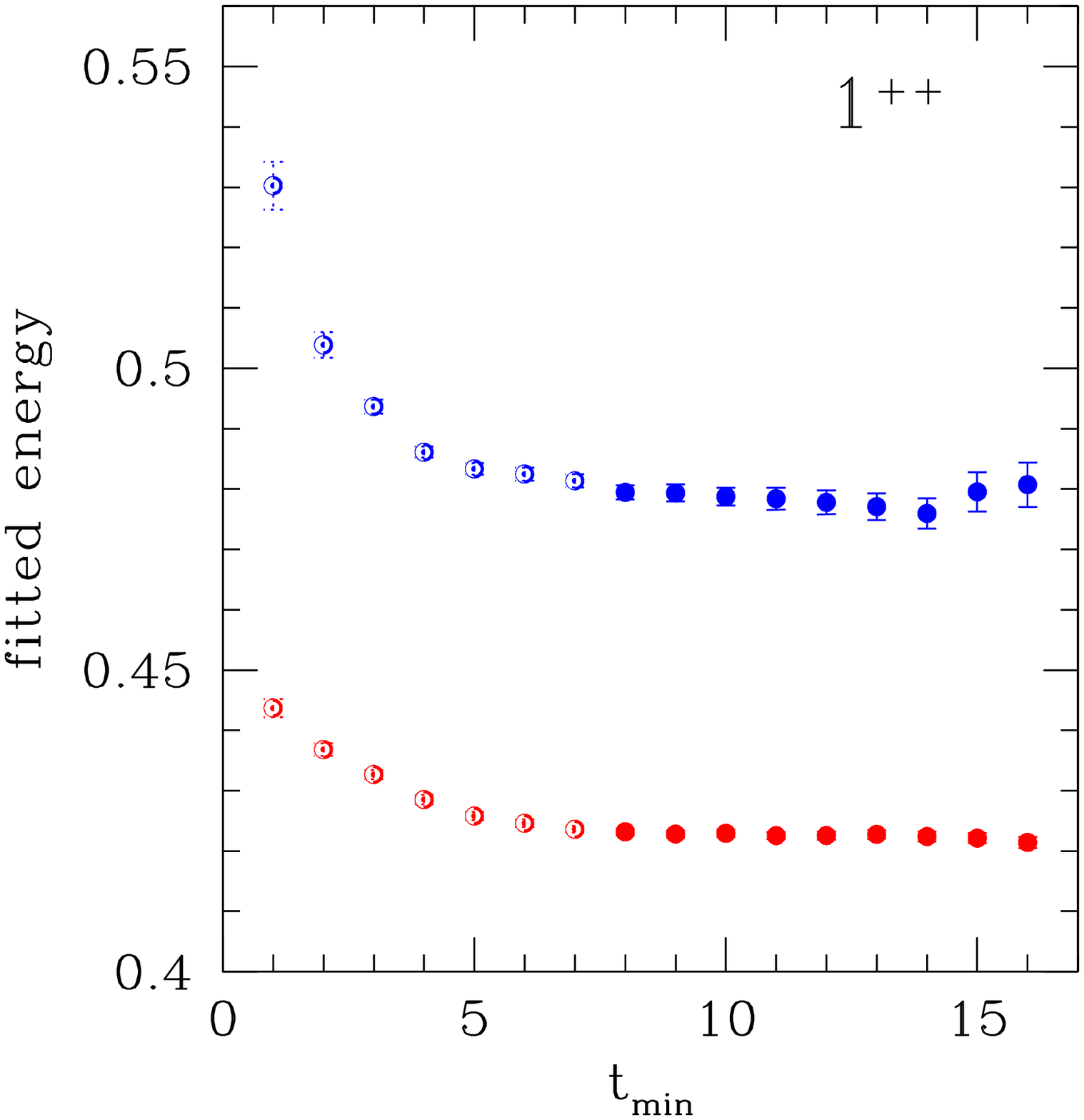}
\vskip -0.3cm
\caption{The effective mass of the $1^{++}$ ground state is shown in the left figure. The ``$\tmin$ plot'' for this state is shown on the right. The open symbols in this plot indicate that the fit did not pass the requirements for a good fit (see text).}
\end{center}
\vskip -0.3cm
\end{figure}

\subsection{Setting the scale}
We estimate the temporal lattice spacing from the S-P splitting in the charmonium sector. The results are tabulated in Table~\ref{table:scale}. We note that the bare quark masses were not quite tuned to the charm quark mass and the puzzling dependence on the mass and/or the effects of the stout-link is still under investigation. 
\begin{table}
\begin{center}
\begin{tabular}{rrrcc}
\hline
 Splitting & stout & $m_0$ & $a_t\Delta$ & $a_t^{-1}\ (\rm{GeV})$  \\ \hline\hline
$\bar{\rm{P}}-\bar{\rm{S}}$ & with & 0.09 & 0.0568(5) & 8.06(7) \\
$\bar{\rm{P}}-\bar{\rm{S}}$ & w/o & 0.12 & 0.0668(7) & 6.85(7) \\\hline
\end{tabular}
\end{center}
\caption{Temporal lattice spacings determined from spin-averaged SP splittings in the charmonium sector for the two actions used in this study.}
\label{table:scale}
\end{table}

\subsection{Dispersion relation}

\begin{figure}[ht]
\begin{center}
\includegraphics[width=2.8in,angle=270]{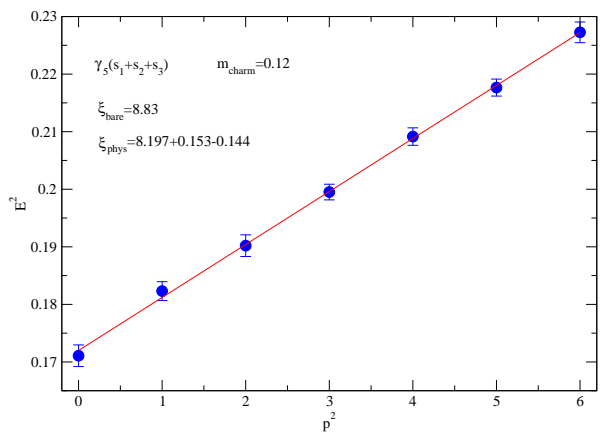}\label{fig:disp}
\vskip -0.3cm
\caption{The dispersion relation computed for $\eta_c$ at a bare quark mass of 0.12. }
\end{center}
\vskip -0.3cm
\end{figure}

In order to check that the $aM$ discretization errors are under
control, we compute the dispersion relation for $\eta_c$ (or $J/\psi$)
at various values of the momenta. The result is shown in
Fig.~\ref{fig:disp}. Since the renormalized anisotropy enters into
this calculation, we cannot at this point determine unambiguously
whether the kinetic mass is equal to the rest mass or not.  This would
require a determination of the dispersion relation of another state,
e.g.\ $D_s$.  We find a renormalized anisotropy determined at the
charm quark mass which is significantly higher than what is found at
the strange quark mass \cite{Morrin:2005xx}.  This discrepancy may be
due to $aM$ errors, but it may also have its origins in the different
methods used to determine the anisotropy (point propagators in
\cite{Morrin:2005xx} vs.\ all-to-all propagators here), combined with
the short temporal extent of our lattice.  This is currently under
investigation, using lattices with a longer time extent.

\section{Results and discussion}

\begin{figure}[ht]
\begin{center}
\includegraphics[width=2.8in]{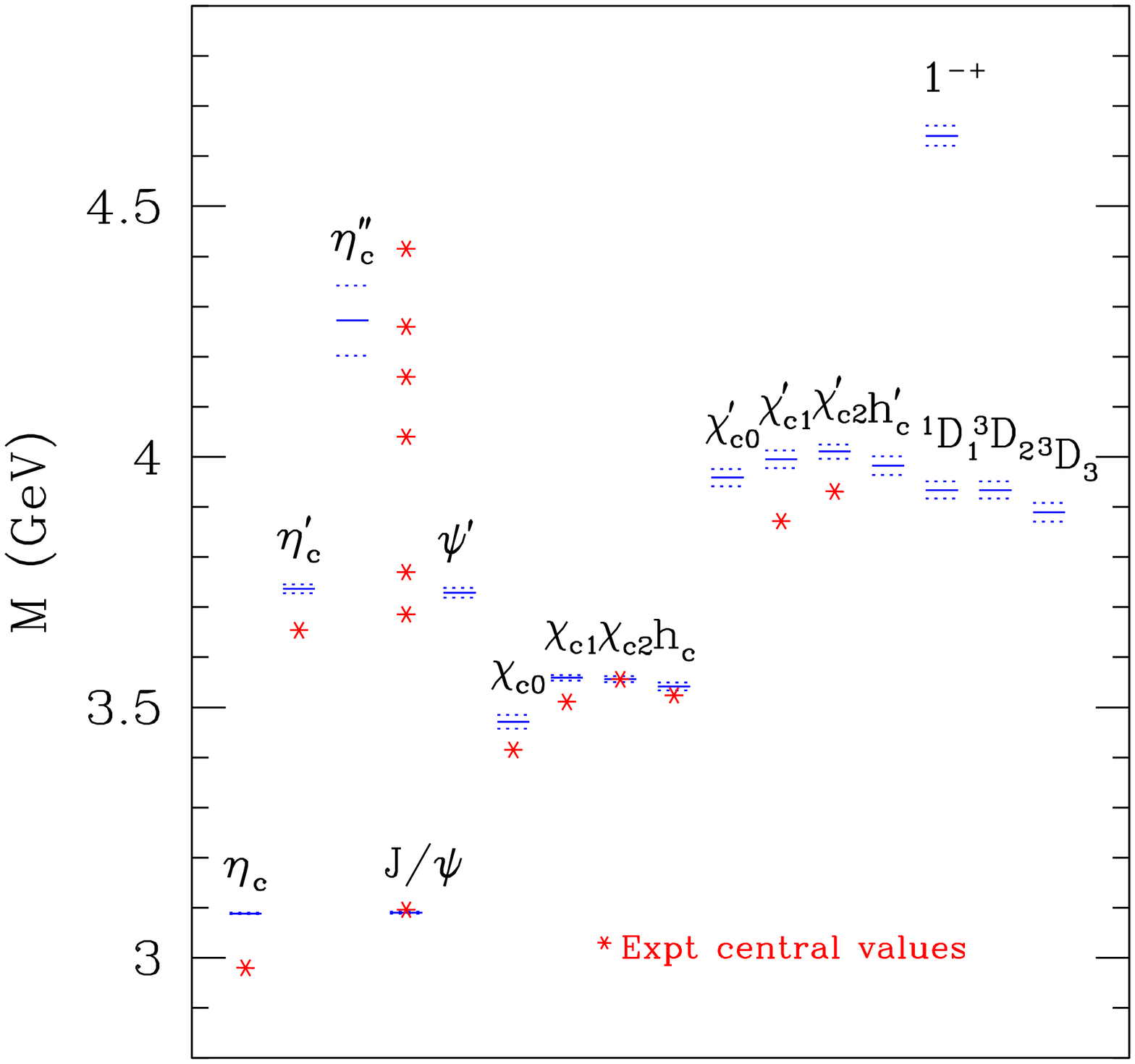}
\includegraphics[width=2.8in]{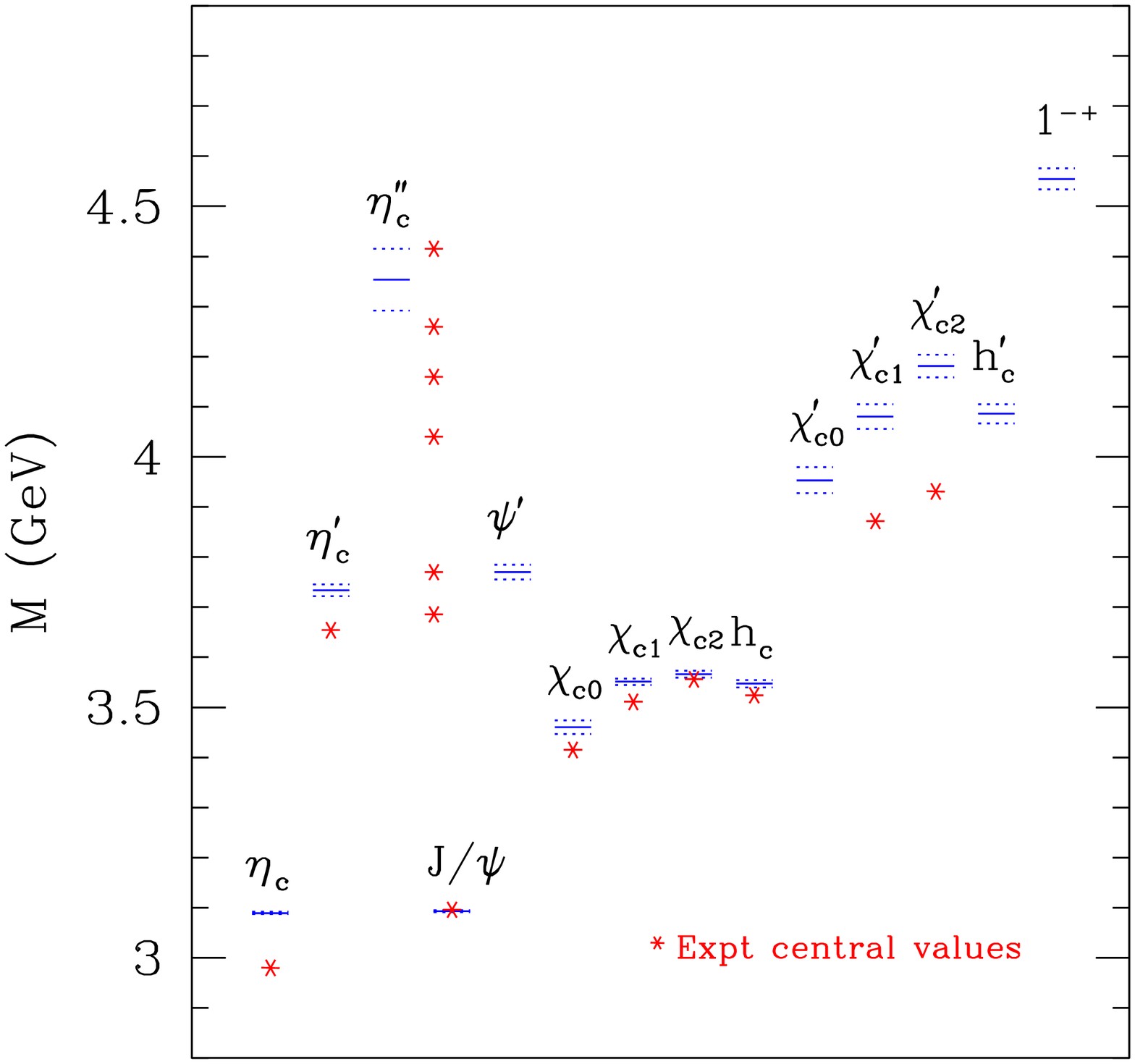}
\vskip -0.3cm
\caption{The charmonium spectrum from the simulation using stout links
  (left) and without stout links (right). The quantum numbers of some
  of the experimental findings for the higher lying P-waves have not
  been established yet; the assignment shown here is the one presented
  in Ref.~\cite{Swanson:2003tb}. }
\end{center}
\vskip -0.3cm
\label{fig:wstout}
\end{figure}

The spectrum computed with (without) stout links is shown on the left
(right) of Fig.~\ref{fig:wstout}. Since the quark mass has not been
exactly tuned to the charm mass, the spectra are shown with the mass
of the $J/\psi$ adjusted to its physical value. The main difference
between the two spectra appears to be a change in the scale. Whether
this is an artifact of the fitting criteria or the diagonalization
procedure, or whether it is a real effect, is under investigation. 

\subsection{Hyperfine splitting} 
All previous lattice studies have given a too small value for the
hyperfine splitting. It is clear that this problem is not yet resolved
with this action. At this stage, the $\eta_c$ and $J/\psi$ are
degenerate within the errors. There are several reasons why this
splitting is not observed:
\begin{itemize}
\item {\bf Finite volume:} The sides of this lattice are roughly 1.6 fm, which is quite small for charmonium.
At this volume, finite temperature effects may also play a role \cite{Morrin:2005zq}.
\item {\bf Higher order improvements of the action:}  The action used
 in this study has leading discretization errors of
 ${\cal O}(a_t,\alpha_sa_s,a_s^3)$.  Higher order correction terms, in
 particular the $\sigma\cdot B$ term, may have a significant impact on
 the hyperfine splitting.
\item {\bf Light sea quark effects:}
The fact that the hyperfine splitting comes out too small in the
quenched approximation indicates that light sea quark effects are important, and the sea quark mass that we are using may still be too large. 
\end{itemize}

\subsection{States above threshold}
We have computed the masses of states which may lie above threshold,
even at the relatively heavy sea quark mass used here. If this is so,
then we must include multi-particle states in the spectrum and perform
a finite volume analysis before we can claim that we have determined
the energy of these states. Multi-particle operators are accessible
with all-to-all quark propagators and investigations in this direction
are under way.

\begin{figure}[ht]
\begin{center}
\includegraphics[width=2.8in]{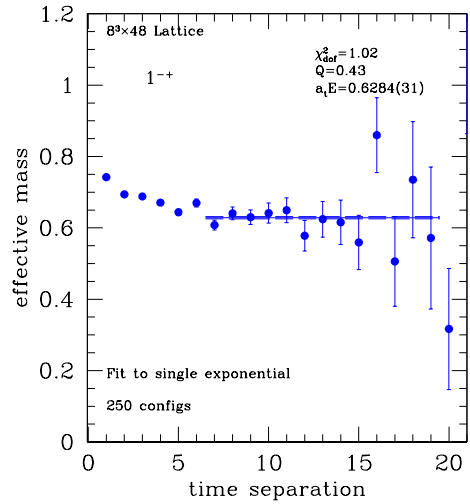}
\vskip -0.3cm
\caption{Effective mass of the $1^{-+}$.\label{fig:exotic}}
\end{center}
\vskip -0.3cm
\end{figure}

Figure \ref{fig:exotic} shows the effective mass for the exotic hybrid
($1^{-+}$), which is found to lie higher than in some quenched
calculations (see \cite{Bernard:1997ib} for example). For this case,
however, it is probably even more important that a larger volume is
used since the wavefunction of the exotic hybrid could be quite large
\cite{Juge:2003qd} and would be squeezed in the small lattices used
here.
\subsection{Hybrid content of $1^{--}$}

\begin{figure}[ht]
\begin{center}
\includegraphics[width=2.8in]{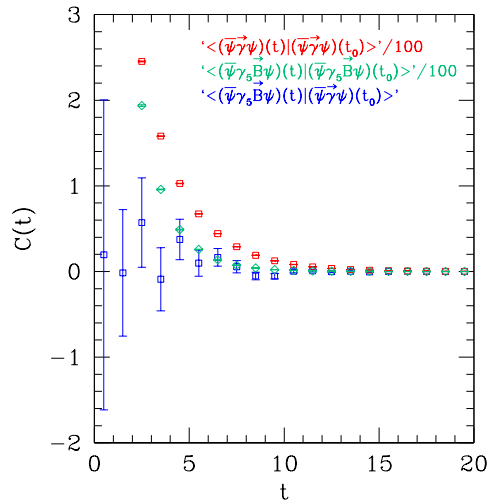}
\vskip -0.3cm
\caption{Correlation functions of the $1^{--}$ state with different interpolating operators.\label{fig:mixing}}
\end{center}
\vskip -0.3cm
\end{figure}

The correlation matrix element in the $1^{--}$ channel between the
state created using a standard $\bar{c}\vec{\gamma}c$ operator and a
hybrid $\bar{c}\gamma_5 \vec{B} c$ operator is plotted in
Fig.~\ref{fig:mixing}. The coefficient appears to be zero within
errors, suggesting that the hybrid content of the $\psi$ spectrum is
small. The hybrid content of the $\Upsilon$ spectrum was also found to
be small in a quenched calculation \cite{Burch:2003zf}. However,
further reduction in noise is required before drawing any definitive
conclusion.

\subsection{Disconnected diagrams}

Disconnected diagrams were not evaluated in this first study as these
contributions are expected to be small (see \cite{deForcrand:2004ia}
for example). However, the inclusion of these diagrams is not
difficult with the all-to-all propagators used here. A complete study
with tuned parameters is planned to study the effect of these
neglected diagrams on hyperfine splittings.

\section{Outlook}

A first study of the charmonium spectrum from an anisotropic, $N_f=2$
action was presented. All-to-all propagators and many operators were
used to optimize the overlap with the standard states and obtain
signals for higher lying radial excitations. A good signal for D-waves
and exotic hybrid states was obtained with a relatively low level of
dilution. However, short distance effects such as the hyperfine
splittings require further investigation. A more detailed study on a
larger volume and with more precisely tuned parameters is in progress.

\section*{Acknowledgements}
The authors would like to thank Jon-Ivar Skullerud for useful discussions. This work has been supported by the SFI grant 04/BRG/P0275 and the IITAC PRTLI initiative. 

\bibliography{trinlat}

\end{document}